\documentclass{appolb}
\usepackage{calc}

\usepackage{graphicx}
\usepackage{dcolumn}
\usepackage{bm}
\usepackage{slashed}
\usepackage{amsmath,graphicx}
\usepackage[colorlinks=true,linktocpage=true,linkcolor=blue,citecolor=blue]{hyperref}
\usepackage{float}
\usepackage{nicefrac}
\usepackage[normalem]{ulem}
\usepackage{amsmath}
\usepackage{subfigure}

\usepackage{float} % for forcing figures locations
\usepackage{bbold}

\newcommand{\beq}{\begin{eqnarray}}
\newcommand{\eeq}{\end{eqnarray}}

\newcommand{\bel}[1]{\begin{eqnarray}\label{#1}}
\newcommand{\eel}{\end{eqnarray}}

\newcommand{\rfn}[1]{(\ref{#1})}

\newcommand{\nn}{\nonumber}

\newcommand{\p}{\partial}

\newcommand{\f}[2]{\frac{#1}{#2}}
\newcommand{\onehalf}{{\nicefrac{1}{2}}}

\renewcommand\sout{\bgroup \color{blue} \ULdepth=-.5ex \ULset}
\newcommand{\ed}{{\varepsilon}}       % energy density
\def\n0{n_{(0)}}
\def\e0{\varepsilon_{(0)}}
\def\P0{P_{(0)}}
\def\s0{s_{(0)}}

\def\umU{u^\mu}

\def\omnL{\omega_{\mu\nu}}
\def\omnU{\omega^{\mu\nu}}

\def\S0iU{{\Sigma}^{0i}}

\def\g5{\gamma_5}

\def\Ot{\tilde \Omega}

\newcommand{\lp}{\left(}
\newcommand{\rp}{\right)}
\newcommand{\lsb}{\left[}
\newcommand{\rsb}{\right]}
\begin{document}
\title{
Dynamics of relativistic spin-polarized fluids
\thanks{Presented by Radoslaw Ryblewski at
                                 XIII Workshop on Particle Correlations and Femtoscopy (WPCF) 2018, May 22-26, 2018, Krak\'{o}w, Poland.}%
}
\author{Wojciech Florkowski
\address{Institute of Nuclear Physics, PL-31342 Krak\'ow, Poland}\\
	 \vspace{0.3cm}
	Bengt Friman
		\address{GSI Helmholtzzentrum f\"{u}r Schwerionenforschung, D-64291 Darmstadt, Germany}	\\
	 \vspace{0.3cm}
		Amaresh Jaiswal
		\address{ 
			School of Physical Sciences, National Institute of Science Education and
			Research, HBNI, Jatni-752050, India}\\
	 \vspace{0.3cm}
		 Radoslaw Ryblewski
		 \address{ 
		 	Institute of Nuclear Physics, PL-31342 Krak\'ow, Poland}
		 \and
			Enrico Speranza
			\address{
				Institute for Theoretical Physics, Goethe University,\\
				D-60-438 Frankfurt am Main, Germany}
}
\maketitle
\begin{abstract}
 We briefly review the foundations of a new relativistic fluid dynamics framework for polarized systems of particles with spin one half. Using this approach we numerically study the dynamics of the spin polarization of a rotating  medium resembling the ones created in high-energy heavy-ion collisions.
\end{abstract}
%
%%%%%%%%%%%%%%%%%%%%%%%%%%%%%%%%%%%%%%%%%%%%%%%%%% 
\section{Introduction}
%%%%%%%%%%%%%%%%%%%%%%%%%%%%%%%%%%%%%%%%%%%%%%%%%%
% 
Very recently the STAR collaboration \cite{STAR:2017ckg} made an intriguing observation of non-zero global spin polarization of $\Lambda$ hyperons emitted from the medium produced in high-energy heavy-ion collisions. This raised the questions concerning the relation between  polarization and vorticity of matter created in these processes. The coupling between the two arises in the global equilibrium state of a rotating system~\cite{Becattini:2009wh,Becattini:2013fla}. However, the most natural framework for such studies is provided by relativistic hydrodynamics, which forms the basis of our current understanding of heavy-ion collisions dynamics  \cite{Florkowski:2010zz,Florkowski:2017olj}. Recently, a new framework for relativistic perfect fluid hydrodynamics of spin-polarized media was presented~\cite{Florkowski:2017ruc,Florkowski:2017dyn} (see also Refs.~\cite{Florkowski:2017njj,Florkowski:2018myy}), aiming at an extension of the work of Refs.~\cite{Becattini:2009wh,Becattini:2013fla} to systems in local equilibrium. In this contribution we review the framework proposed in Refs.~\cite{Florkowski:2017ruc,Florkowski:2017dyn,Florkowski:2017njj,Florkowski:2018myy} and use it for numerical studies of polarization dynamics in heavy-ion collisions.
%
%%%%%%%%%%%%%%%%%%%%%%%%%%%%%%%%%%%%%%%%%%%%%%%%%%
\section{Evolution equations for spin-polarized fluids}
%%%%%%%%%%%%%%%%%%%%%%%%%%%%%%%%%%%%%%%%%%%%%%%%%%
%
In Refs.~\cite{Florkowski:2017ruc,Florkowski:2017dyn} a fluid dynamical framework for polarized systems of \mbox{spin-$\onehalf$} particles and antiparticles was proposed. It is based on conservation laws of baryon charge, energy, linear momentum and total angular momentum 
\beq
\p_\alpha N^{\alpha} &=& 0,\label{Ncons}\\
\p_\alpha T^{\alpha\beta} &=& 0,\label{Tcons} \\
\p_\alpha J^{\alpha,\beta\gamma} &=& 0.\label{Jcons}
\eeq
Employing the kinetic-theory definitions of Ref.~\cite{DeGroot:1980dk}
together with the equilibrium spin density matrices proposed in Ref.~\cite{Becattini:2013fla}, one finds that the baryon current $N^{\alpha}$ and the energy-momentum tensor $T^{\alpha\beta}$ in Eqs.~(\ref{Ncons})-(\ref{Tcons}) have the following perfect-fluid structure 
\beq 
N^{\alpha} &=&  n u^\alpha, \label{N}\\
T^{\alpha \beta} &=&   (\ed + P) u^\alpha u^\beta - P g^{\alpha \beta}, \label{T} 
\eeq
respectively, where $g^{\alpha \beta}={\rm diag}(+1,-1,-1,-1)$ is the metric tensor and $u^\alpha$ denotes the four-velocity of the fluid. In addition, one finds that the entropy current, computed with the Boltzmann formula, takes the form 
\beq 
S^{\alpha} &=&  s u^\alpha.\label{entropycurrent} 
\eeq
Rewriting the total angular momentum tensor $J^{\alpha,\beta\gamma}$ in Eq.~(\ref{Jcons}) as a sum of the orbital $L^{\alpha,\beta\gamma} = x^{\beta} T^{\gamma\alpha} - x^{\gamma} T^{\beta\alpha}$ and spin $S^{\alpha,\beta\gamma}$ parts and employing Eqs.~(\ref{Tcons}) and (\ref{T}) one finds that the spin tensor is separately conserved, $\p_\alpha S^{\alpha,\beta\gamma} = 0$. Moreover, assuming that the latter has the form proposed in Ref.~\cite{Becattini:2009wh}, namely,
\beq 
S^{\alpha,\beta \gamma}= \frac{w u^\alpha}{4 \zeta} \omega^{\beta\gamma}, \label{S} 
\eeq
Eq.~(\ref{Jcons}) takes the form
\beq
\dot{\bar{\omega}}^{\mu\nu}   &=&0 \label{dotomegabar}.  
\eeq
Herein,  $\omnU$ is the spin polarization tensor, which is antisymmetric and satisfies the relation  $\epsilon_{\alpha\beta\gamma\delta} \omega^{\alpha\beta}\omega^{\gamma\delta}=0$, $\bar{\omega}^{\mu\nu}\equiv\omega^{\mu\nu}/(2\zeta)$ is the normalized spin polarization tensor with $\zeta  \equiv \f{1}{2 \sqrt{2}} \sqrt{ \omnL \omnU}$ being real,  and $\dot{(\hphantom{A})}\equiv u \cdot\p$ denotes the comoving derivative. 

The thermodynamic quantities, namely, energy density, pressure, baryon density, and spin density, 
\beq
&&\ed = 4  \cosh(\zeta)  \cosh(\xi)  \e0(T) \label{edensity}, \qquad
P = 4  \cosh(\zeta)  \cosh(\xi) \P0(T) \label{pressure},\\
&& n = 4 \cosh(\zeta)   \sinh(\xi)  \n0(T) \label{bdensity},\qquad
w = 4  \sinh(\zeta) \cosh(\xi) \n0(T) \label{sdensity},
\eeq
respectively, satisfy the fundamental
thermodynamic relation $\ed + P= s T +\mu n + \Omega w$ with $s = 4  \cosh(\zeta)  \cosh(\xi) \s0(T)$ being the entropy density. The quantities $\xi \equiv \mu/T$ and $\zeta \equiv \Omega/T$ parametrize the baryon $\mu$ and the spin $\Omega$ chemical potentials, and are, together with the temperature $T$, treated as the independent thermodynamic variables entering the grand canonical potential. The thermodynamic quantities Eqs.~(\ref{edensity})--(\ref{sdensity}) are expressed in terms of auxiliary ones describing a corresponding system of spin-$0$ particles
\beq
\n0(T) &=&  \f{\kappa}{2\pi^2}\, T^3 \, \hat{m}^2 K_2\left( \hat{m}\right) \,, \label{polden}\nn\\
\e0(T) &=&   \f{\kappa}{2\pi^2} \, T^4 \, \hat{m} ^2
\Big[ 3 K_{2}\left( \hat{m} \right) + \hat{m}  K_{1} \left( \hat{m}  \right) \Big]\,,  \label{eneden}\nn\\
\P0(T) &=&  T \, \n0(T)  \,, \label{P0}\nn
\eeq
where $\s0(T) = \f{1}{T} \lsb\e0(T)+\P0(T)\rsb$, $\hat{m}\equiv m/T$ and $\kappa \equiv g/(2\pi)^3$, with $g$ denoting the number of internal degrees of freedom excluding spin.

It is instructive to study projections of Eqs.~(\ref{Tcons}). In particular, one finds that the projection of the energy-momentum conservation law onto directions orthogonal to the fluid four-velocity leads to the relativistic Euler equations 
\beq
(\ed + P)  \dot{u}^\mu &=& \p^\mu P - \umU \dot{P}\label{emcprojt}\,,
\eeq
where  $\theta\equiv \p\cdot u$ is the expansion scalar. On the other hand, projecting Eq.~(\ref{Tcons}) onto the fluid flow and using the differentials of the pressure $P=P(T,\mu,\Omega)$ yields
\bel{emcuproj}
T \p_\mu (s \umU)+\mu\, \p_\mu (n \umU)+\Omega\, \p_\mu (w \umU)=0. 
\eel
Requiring entropy and baryon number conservation, which makes first and second term in (\ref{emcuproj}) vanish, leads to
\beq
 \p_\mu (s \umU)=\dot{s} + s\, \theta &=&0 \label{entcons}\,, \\
 \p_\mu (n \umU)=\dot{n} + n\, \theta&=&0\label{chargcons}\,, \\
 \p_\mu (w \umU)=\dot{w} + w\, \theta&=&0\label{polcons}\,.
\eeq
Equations \rfn{emcprojt}, \rfn{entcons}, \rfn{chargcons} and \rfn{polcons} are six coupled partial differential equations which determine the dynamics of the space-time dependent quantities $\mu(x)$, $\Omega(x) \equiv \f{T}{2 \sqrt{2}} \sqrt{ \omnL \omnU}$, $T(x)$ and $u^\mu(x)$. On top of their evolution one has to solve Eqs.~(\ref{dotomegabar}) for the normalized components of the polarization tensor.

In Refs.~\cite{Florkowski:2017ruc,Florkowski:2017dyn} it was shown that Eqs.~\rfn{emcprojt}, \rfn{entcons}, \rfn{chargcons} and \rfn{polcons}  have the stationary vortex-like solution corresponding to a rotating global equilibrium state \cite{Becattini:2009wh,Becattini:2013fla} with 
\beq
u^\mu = \gamma \,(1,  -  \Ot \, y,    \, \Ot \, x,  0), \label{uvortex}
\eeq
$T = T_0 \gamma$,  $\mu = \mu_0 \gamma$ and  $\Omega = \Omega_0 \gamma$ 
where  $\gamma = 1/\sqrt{1 - \Ot^2 r^2}$ is the Lorentz factor,   $r = \sqrt{x^2 + y^2}$  and $T_0$, $\mu_0$, and $\Omega_0$ are arbitrary constants. The corresponding nontrivial form of the polarization tensor is  ${\omega}_{ij}=-{  \omega}_{ji} = \Ot/T_0  = 2 \, \Omega_0 /T_0$ for $i=x$ and $j=y$ and ${\omega}_{ij}=0$ otherwise. In the next section we study Eqs.~\rfn{emcprojt}, \rfn{entcons}, \rfn{chargcons}, \rfn{polcons} and \rfn{dotomegabar} in the case where the equilibrium is achieved only locally.
%
%%%%%%%%%%%%%%%%%%%%%%%%%%%%%%%%%%%%%%%%%%%%%%%%%%
\section{Numerical results}
\begin{figure}[t]
 \begin{center}
     \includegraphics[angle=0,width=0.9 \textwidth]{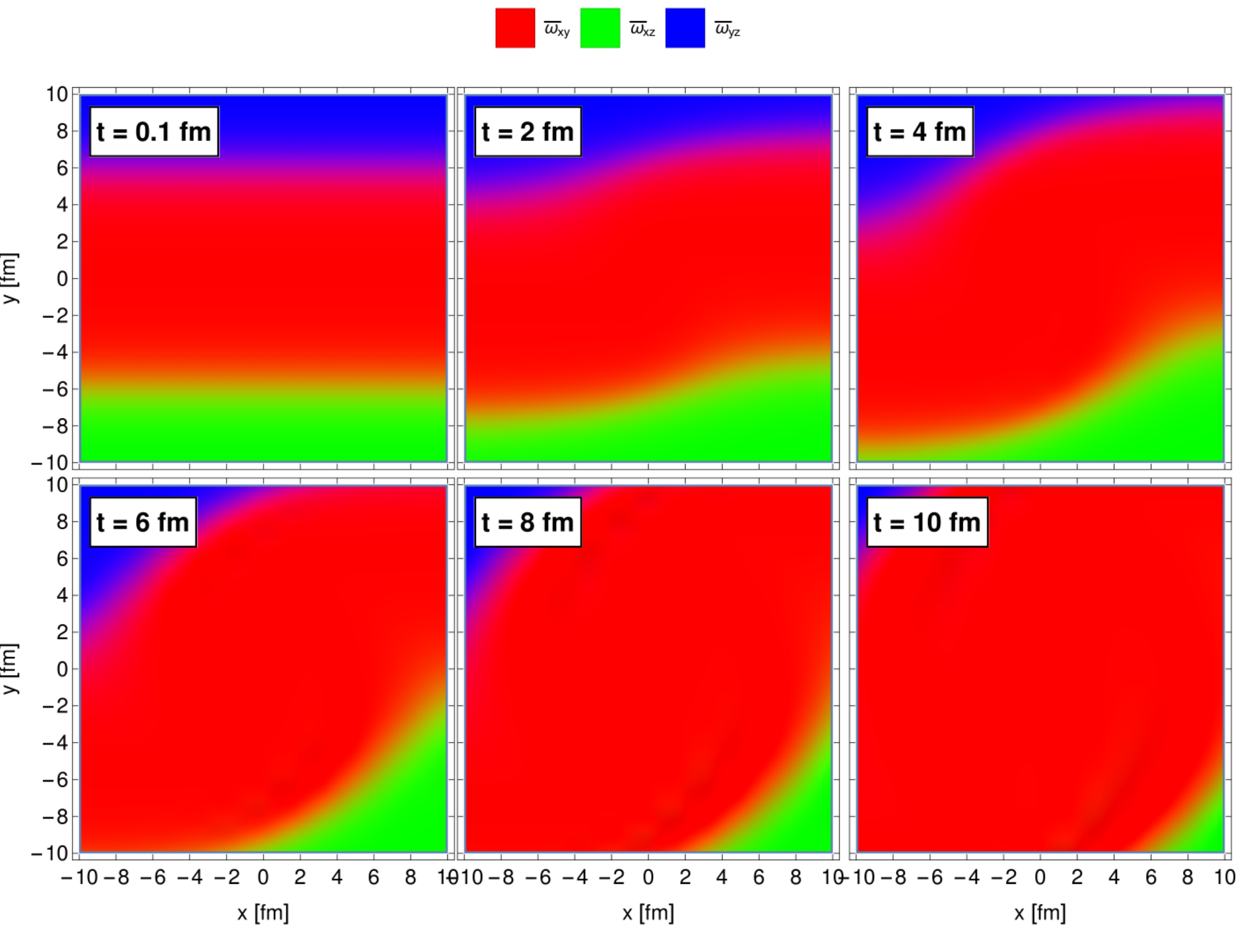}
 \end{center}
 	\caption{(Color online) The components of the spin polarization tensor: ${\bar \omega}_{xy}$ (red), ${\bar \omega}_{xz}$ (green) and ${\bar \omega}_{yz}$ (blue) in the reaction plane ($x-y$) at times: $t=0.1, 2, 4, 6, 8, 10$ fm.
 	}
 	\label{fig:sol}
 \end{figure}

%%%%%%%%%%%%%%%%%%%%%%%%%%%%%%%%%%%%%%%%%%%%%%%%%%
%
In the following we study numerically the solutions of the hydrodynamic equations presented above for a rotating spin-polarized medium, which resembles the systems created in the low-energy heavy-ion collisions. It is modelled by a gaussian source $T_{\rm i} = T_0 \,\,g(x,y,z)$
where $g(x,y,z)=\exp\lp -\frac{x^2}{2 \sigma_x^2}-\frac{y^2}{2 \sigma_y ^2}-\frac{z^2}{2 \sigma_z ^2}\rp$. The initial flow is assumed to have the form \rfn{uvortex}, where we replace $\Ot$ with $\lp 1/r  \rp\tanh\lp r/r_0 \rp$. The parameter $r_0=1$ sets the rotation speed. The initial spin chemical potential is given by $\Omega_{\rm i} = 0.03 \,T_{\rm i}/2$,  while the initial baryon chemical potential is given by $\mu_{\rm i} = \mu_0 \,\,g(x,y,z)$, where $\mu_0=200$ MeV.
With this setup we let the system evolve in Minkowski time starting at $t_0=0.1$ fm using Eqs.~\rfn{emcprojt}, \rfn{entcons}, \rfn{chargcons} and \rfn{polcons}.  Once the evolution is finished we initialize the normalized polarization tensor with
\bel{omB}
{\bar \omega}_{\mu\nu}= 
\begin{bmatrix}
	0       &  0 & 0 & 0 \\
	0  &  0    & {\bar \omega}_{xy} & {\bar \omega}_{xz} \\
	0  & -{\bar \omega}_{xy} & 0 & {\bar \omega}_{yz} \\
	0  & -{\bar \omega}_{xz} & -{\bar \omega}_{yz} & 0
\end{bmatrix},
\eel
and evolve it on top of our hydrodynamic results, using  Eqs.~(\ref{dotomegabar}) starting from $t=t_0$. The results of the latter calculations are presented in Fig.~\ref{fig:sol}. Initially (top left panel) the ${\bar \omega}_{xy}$ (red) component dominate $|y| < 6$ region, while the  ${\bar \omega}_{yz}$ and  ${\bar \omega}_{zz}$ components dominate the $y > 6$ and  $y < -6$ regions, respectively. During the subsequent evolution, the spin polarization is transferred to different regions of space due to the non-trivial velocity field of the medium. At the very end of the evolution the component ${\bar \omega}_{xy}$ dominates over the other components in almost the entire space.

Our study shows that, if the dynamics of the spin polarization is included in the fluid modelling, the spin-polarization of the medium may change significantly over its evolution. Such a fluid dynamic stage is missing in the models used so far for interpreting the data.  Thus, it is of great importance to properly model the early-time dynamics of the spin polarization as this may significantly affect the final results.
 % 
 %%%%%%%%%%%%%%%%%%%%%%
 %%%%%%%%%%%%%%%%%%%%%%
 % 
%%%%%%%%%%%%%%%%%%%%%%%%%%%%%%%%%%%%%%%%%%%%%%%%%%
\section{Summary}
%%%%%%%%%%%%%%%%%%%%%%%%%%%%%%%%%%%%%%%%%%%%%%%%%%
%
Employing a novel formulation of relativistic perfect fluid hydrodynamics for polarized systems of particles with spin $\onehalf$ we numerically studied the evolution of a rotating spin-polarized source resembling the ones created in high-energy heavy-ion collisions. We find that spin polarization tensor undergoes a non-trivial evolution in the hydrodynamic stage which may significantly affect the final results used for the  interpretation of experimental data.
%
%%%%%%%%%%%%%%%%%%%%%%%%%%%%%%%%%%%%%%%%%%%%%%%%%%
\section*{Acknowledgments}
%%%%%%%%%%%%%%%%%%%%%%%%%%%%%%%%%%%%%%%%%%%%%%%%%%
%
W.F. and R.R were supported in part by the Polish National Science Center Grant No.   2016/23/B/ST2/00717.  E.S.  was  supported  by  BMBF  Verbundprojekt 05P2015 - Alice at High Rate.  E.S. acknowledges partial support by the  Deutsche  Forschungsgemeinschaft  (DFG)  through  the  grant  CRC-TR 211 “Strong-interaction matter under extreme conditions”. A.J. is supported in part by the DST-INSPIRE faculty award under Grant No. DST/INSPIRE/04/2017/000038. This research was supported in part by the ExtreMe Matter Institute EMMI at GSI and was performed in the framework of COST Action CA15213 “Theory of hot matter and relativistic heavy-ion collisions” (THOR).
%
%%%%%%%%%%%%%%%%%%%%%%%%%%%%%%%%%%%%%%%%%%%%%%%%%%


\begin{thebibliography}{100}
\expandafter\ifx\csname url\endcsname\relax \def\url#1{{\tt #1}}\fi
\expandafter\ifx\csname urlprefix\endcsname\relax\def\urlprefix{URL}\fi
\providecommand{\eprint}[2][]{\url{#2}}

%\cite{STAR:2017ckg}
\bibitem{STAR:2017ckg} 
L.~Adamczyk {\it et al.} [STAR Collaboration],
%``Global $\Lambda$ hyperon polarization in nuclear collisions: evidence for the most vortical fluid,''
Nature {\bf 548}, 62 (2017)
%doi:10.1038/nature23004
%[arXiv:1701.06657 [nucl-ex]].
%%CITATION = doi:10.1038/nature23004;%%
%68 citations counted in INSPIRE as of 01 Jun 2018


%\cite{Becattini:2009wh}
\bibitem{Becattini:2009wh} 
F.~Becattini and L.~Tinti,
%``The Ideal relativistic rotating gas as a perfect fluid with spin,''
Annals Phys.\  {\bf 325}, 1566 (2010)
%doi:10.1016/j.aop.2010.03.007
%[arXiv:0911.0864 [gr-qc]].
%%CITATION = doi:10.1016/j.aop.2010.03.007;%%
%25 citations counted in INSPIRE as of 01 Jun 2018

 %\cite{Becattini:2013fla}
 \bibitem{Becattini:2013fla} 
 F.~Becattini, V.~Chandra, L.~Del Zanna and E.~Grossi,
 %``Relativistic distribution function for particles with spin at local thermodynamical equilibrium,''
 Annals Phys.\  {\bf 338}, 32 (2013)
 %doi:10.1016/j.aop.2013.07.004
 %[arXiv:1303.3431 [nucl-th]].
 %%CITATION = doi:10.1016/j.aop.2013.07.004;%%
 %57 citations counted in INSPIRE as of 01 Jun 2018 
 
%\cite{Florkowski:2010zz}
\bibitem{Florkowski:2010zz} 
  W.~Florkowski,
  ``Phenomenology of Ultra-Relativistic Heavy-Ion Collisions,''
  Singapore, Singapore: World Scientific (2010) 416 p
  %73 citations counted in INSPIRE as of 13 Jul 2017
 
 %\cite{Florkowski:2017olj}
 \bibitem{Florkowski:2017olj} 
 W.~Florkowski, M.~P.~Heller and M.~Spalinski,
 %``New theories of relativistic hydrodynamics in the LHC era,''
 Rept.\ Prog.\ Phys.\  {\bf 81}, no. 4, 046001 (2018)
% doi:10.1088/1361-6633/aaa091
% [arXiv:1707.02282 [hep-ph]].
 %%CITATION = doi:10.1088/1361-6633/aaa091;%%
 %39 citations counted in INSPIRE as of 01 Jun 2018
 

 
%\cite{Florkowski:2017ruc}
\bibitem{Florkowski:2017ruc} 
W.~Florkowski, B.~Friman, A.~Jaiswal and E.~Speranza,
%``Relativistic fluid dynamics with spin,''
Phys.\ Rev.\ C {\bf 97}, no. 4, 041901 (2018)
%doi:10.1103/PhysRevC.97.041901
%[arXiv:1705.00587 [nucl-th]].
%%CITATION = doi:10.1103/PhysRevC.97.041901;%%
%8 citations counted in INSPIRE as of 01 Jun 2018  

%\cite{Florkowski:2017dyn}
\bibitem{Florkowski:2017dyn} 
W.~Florkowski, B.~Friman, A.~Jaiswal, R.~Ryblewski and E.~Speranza,
%``Spin-dependent distribution functions for relativistic hydrodynamics of spin-1/2 particles,''
Phys.\ Rev.\ D {\bf 97}, no. 11, 116017 (2018)
%doi:10.1103/PhysRevD.97.116017
%[arXiv:1712.07676 [nucl-th]].
%%CITATION = doi:10.1103/PhysRevD.97.116017;%%
%7 citations counted in INSPIRE as of 10 Oct 2018


%\cite{Florkowski:2017njj}
\bibitem{Florkowski:2017njj} 
W.~Florkowski, B.~Friman, A.~Jaiswal and E.~Speranza,
%``Relativistic hydrodynamics of particles with spin 1/2,''
Acta Phys.\ Polon.\ Supp.\  {\bf 10}, 1139 (2017)
%doi:10.5506/APhysPolBSupp.10.1139
%[arXiv:1708.04035 [hep-ph]].
%%CITATION = doi:10.5506/APhysPolBSupp.10.1139;%%
%1 citations counted in INSPIRE as of 01 Jun 2018

%\cite{Florkowski:2018myy}
\bibitem{Florkowski:2018myy} 
W.~Florkowski, E.~Speranza and F.~Becattini,
%``Perfect-fluid hydrodynamics with constant acceleration along the stream lines and spin polarization,''
arXiv:1803.11098 [nucl-th].
%%CITATION = ARXIV:1803.11098;%%
%1 citations counted in INSPIRE as of 01 Jun 2018
  


%\cite{DeGroot:1980dk}
\bibitem{DeGroot:1980dk} 
S.~R.~De Groot, W.~A.~Van Leeuwen and C.~G.~Van Weert,
%``Relativistic Kinetic Theory. Principles and Applications,''
Amsterdam, Netherlands: North-holland ( 1980) 417p
%73 citations counted in INSPIRE as of 01 Jun 2018

 
\end{thebibliography}
\end{document}